\documentstyle[psfig]{l-aa}
\begin{document}

   \thesaurus{06         % A&A Section 6: Form. struct. and evolut. of stars
              (02.01.2;  % Accretion, accretion disks,
               02.12.1;  % line: formation,
               02.12.3;  % line: profiles,
               08.14.2;  % novae, cataclysmic variables,
               08.09.2)} % stars: individual: U Gem.

\title{Spiral shocks in the accretion disk of the dwarf nova U Geminorum}
\author{V.V. Neustroev \inst{1,2} and N.V. Borisov \inst{2}}

\offprints{V.V. Neustroev (first address)}

\institute{
           Department of Astronomy and Mechanics, Udmurt State University,
           1, Universitetskaia, Izhevsk, 426034, Russia \\
           e-mail: benj@uni.udm.ru
 \and
           Special Astrophysical Observatory, Nizhnij Arkhyz,
           Karachaevo-Cherkesia, 357147, Russia. E-mail: borisov@sao.ru
}

\date{Received , accepted }

\maketitle

\begin{abstract}
We present a study of the observable spectral manifestations associated with
spiral shocks within quiescent accretion disks and compare them with
observations of the dwarf nova U Geminorum. Our results indicate that the
orbital behaviour of the hydrogen emission line profile in the spectrum of U
Gem can be reproduced via a simple model of the accretion disk with two
symmetrically located spiral shocks with the spiral angle $\theta $ about 60$%
^{\circ }$. This provides for the first time convincing evidence for the
existence of spiral shocks in the accretion disk of a cataclysmic
variable in quiescence.


\end{abstract}

\section{Introduction}

Although the understanding of accretion disk physics has made much progress
in recent years, a viable angular momentum transfer mechanism still remains the
main unresolved problem. At present there are two main standpoints on this
problem. According to first one, angular momentum is transported due to
the presence of
turbulent or magnetic viscosity in the disk (Shakura \& Sunyaev 1973). On
the other hand, hydrodynamical numerical calculations have shown that tidal
forces from the secondary will induce spiral shock waves in the accretion
disk (Sawada et al. 1986), which may provide an efficient transfer
mechanism. It should be noted, that these two mechanisms are mutually
exclusive, as shock waves in the presence of viscosity will be smeared out
(Bunk et al. 1990; Chakrabarti 1990).

\begin{figure}
%\picplace{10 cm}

{\vbox{
\psfig{figure=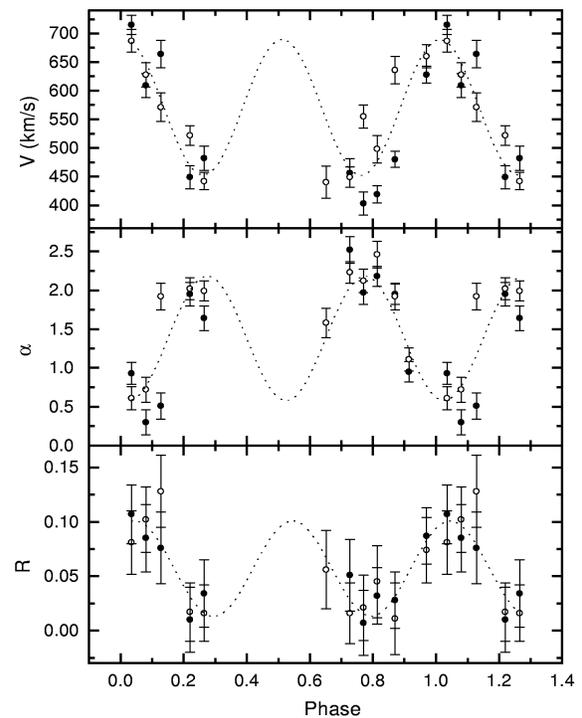,width=8cm,bbllx=0pt,bblly=30pt,bburx=370pt,bbury=500pt,clip=}}
}
\par
%\vspace{-1 cm}

\caption{The orbital phase dependence of the accretion disk parameters,
obtained from modelling H$_{\beta }$ (closed circles) and H$_{\gamma }$
(open circles) emission lines (for details see Borisov \& Neustroev, 1998a)}
\label{fig1}
\end{figure}

Up till now, there were no proof for spiral structure within quiescent
accretion disk. Only Steeghs et al. (1997) have found evidence for spiral
structure in the accretion disk of the dwarf nova IP Pegasi, observed during
outburst.

\begin{figure*}
%\picplace{7.5 cm}

{\vbox{
\psfig{figure=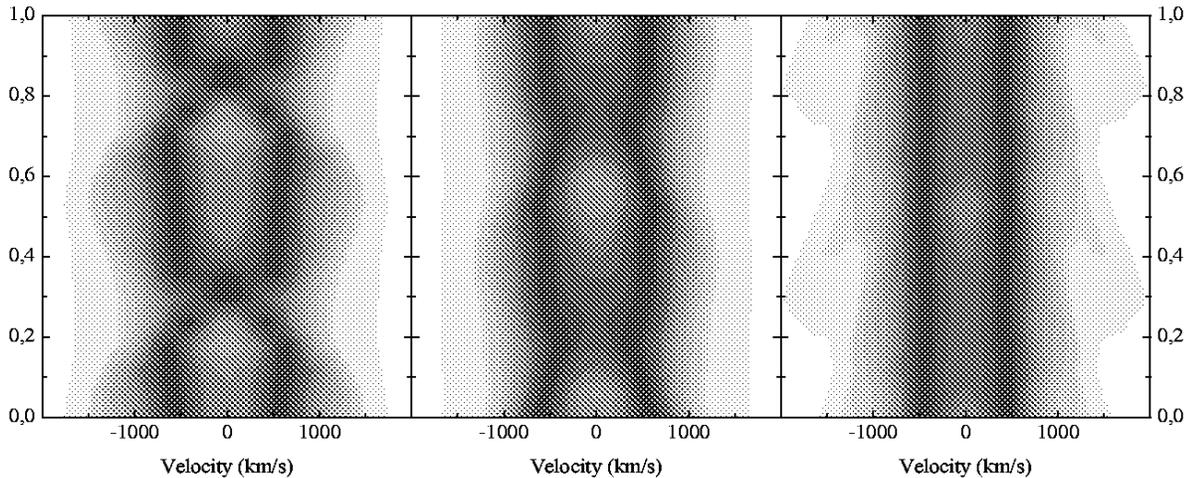,width=17cm,bbllx=40pt,bblly=350pt,bburx=375pt,bbury=500pt,clip=}}%
}
\par
\vspace{-1 cm}

\caption{The predicted trailed spectra of the emission lines arising in
the accretion disk with 2 spiral shocks. The spiral angle $\theta$ is
40$^{\circ }$ (left), 60$^{\circ }$ (middle) and 80$^{\circ }$ (right)}
\label{fig2}
\end{figure*}

The importance regarding the determination of this angular momentum transfer
mechanism for accretion disk theory determines rigourous observational efforts
to confirm such spiral shock phenomena. Preliminary work by Bunk et al.
(1990) investigated the distribution of energy in a continuum spectrum, and
the eclipsing light curve in the presence of spiral shocks on the disk. It
has been shown that the direct photometric detection of spiral shocks is
problematic.

On the other hand, Chakrabarti \& Wiita (1993) have shown that the spiral
shocks in an accretion disk could have significant effects upon observable
emission lines, namely that the double peak separation in emission lines
will vary with the binary's phase. In our recent study of the accretion disk
structure of the dwarf nova U Geminorum in quiescence (Borisov \& Neustroev
1998a), we have
detected a change of the double peak separation in all investigated emission
lines (H$_{\beta }$, H$_{\gamma }$ and H$_{\delta }$) with amplitudes 136,
108 and 60 km/s respectively in agreement with this hypothesis. It is
important to note, that such changes consistent with the shock model were
observed during two separate observation runs.

In this paper we present a more detailed study of the observable spectral
manifestations of spiral shocks than was previously done (Chakrabarti \&
Wiita 1993) and we compare our
conclusions with observational data of the U Gem system.

\section{U Gem: observations and results}

U Gem is one of the best studied cataclysmic variables, being the
prototype of the dwarf nova class. 
%The photometric study of U Gem by
%Krzeminski (1965) showed that U Gem is an eclipsing binary system with an
%orbital period of 4$^{h}15^{m}.$ 
The optical spectrum of U Gem types are
dominated by broad double-peaked emission lines of H\thinspace I,
He\thinspace I and Ca\thinspace II, having their origin in the accretion
disk. 
%Spectra of U Gem also show narrow emission line components which vary
%sinusoidally in wavelength (the so-called ''S-wave'' components), and which
%have previously been ascribed to emission from the gas stream or the bright
%spot.

The spectral data of U Gem we present here were obtained during
January 21, 1994 and December 5-6, 1994 with the 1000-channel television
scanner and the spectrograph SP-124 of the 6-m telescope of the SAO. 
We would like to point out that our January observations were obtained 
after U Gem returned to the minimum light. This happened 30 days after U
Gem's outburst began  which finished 7 days prior to our observations i.e.
around January 14th. December data was collected after the outburst has
finished, which was about 85 days before the observations.
After analysis and modelling of the emission lines we obtained all the basic
parameters of the accretion disk and the bright spot (see Borisov \& Neustroev 1998a).

In Fig.~\ref{fig1} we show the orbital phase dependence of the disk
parameters. Here, R is the ratio of the inner and the outer radius of the
disk, V is the projection of velocity of outer rim of accretion disk on the
line of the sight, $\alpha$ is the emissivity parameter (the line
surface brightness of the disk is assumed to scale as R$^{-\alpha}$). For
details see Borisov \& Neustroev (1998a,b).
In case of moving substance in the disk on keplerian circular
orbits, the peak separation in emission lines (which mostly affects on the
definition of parameter V) should remain constant during an orbital period.
However, it can be seen that all parameters, including V, vary sinusoidally
as $\sin 2\varphi $ (the hypothesis of constant parameters can be rejected
with a confidence probability of more than 99\%). Moreover, the explicit
correlation between change of V, R and $\alpha $ is observed. The multiple
correlation coefficient between V and parameters R and $\alpha $ for all
lines is more than 0.72 with a confidence probability more than 96\%. Such
variability of parameters is present in both data sets and the phases of
their modification are almost identical. 
Please note, that the marked variations of parameters
were not revealed on the other objects observed with
the same devices and in similar conditions (for example, 
WZ Sge - Neustroev 1998).
Of particular interest is the
behaviour of the parameter V as it's variation with the binary's phase
reveals a deviation of the velocity field in the disk from the standard
circular Keplerian model. The possible cause of such a deviation is the
non-circular form of the outer edge of the accretion disk.

\begin{figure}
%\picplace{9 cm}

{\vbox{
\psfig{figure=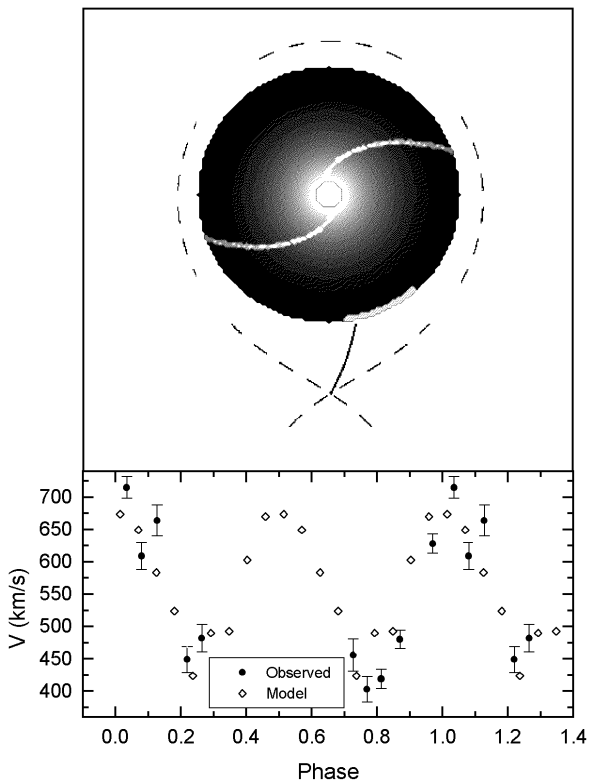,width=8cm,bbllx=0pt,bblly=40pt,bburx=300pt,bbury=440pt,clip=}}}
%\par
%\vspace{-1 cm}

\caption{Possible spatial image of the accretion disk emissivity model (top)
and the predicted orbital phase dependence of the double peak separation
along with actual observed data (bottom)}
\label{fig3}
\end{figure}

Research into the dynamics and stabilities of Keplerian elliptical accretion
disks has shown (Lyubarskij et al. 1994), that such disks
should be stable formations. However, modelling has indicated that the peak
separation in emission lines, formed in the Keplerian elliptic accretion
disk, will remain constant during an orbital period. Therefore, to sustain
the Keplerian velocity field hypothesis in the disk, the phase variability
of the peak separation can only be explained by a deviation in the emission
line source region from a circle in the circular disk. For the elliptic disk
the source region should have at least an eccentricity different from the
disk.

Paczynski (1977) investigated the tidal effect of the secondary star upon
the disk by integrating the orbits of test particles in the restricted
three-body problem to determine the departures from a Keplerian velocity
field in a pressureless disk. He found an elliptical distortion of the outer
disk. The peak separation in emission lines, formed in such a disk, will
vary as
$$
V=V_{0}-\Delta V\cos 2\varphi \eqno (1)
$$
\noindent where $\varphi $ is a phase of an orbital period.{\bf \ }The
dependence of parameters which is obtained (Fig.~\ref{fig1}), however, does
not agree with Eq. (1) because of a shift of phases which is about 90$^{\circ }$%
, and therefore cannot be explained by the model of Paczynski.

Such variations of the peak separation in lines could be caused by spiral
shock in the accretion disk, obtained in many numerical hydrodynamic
calculations (for example, Sawada et al. 1986). An earlier analysis of
the influence of spiral shock on emission lines (Chakrabarti \& Wiita
1993) has shown that the peak separation in such lines under certain
conditions should vary with orbital phase.

\section{Modelling}

Emission line profiles arising in the accretion disk with or without spiral
shocks can be calculated theoretically and by comparing them, the observed
effects of spiral shocks can be derived. For this purpose we used the
following technique. We decided to apply a model which includes a flat
Keplerian geometrically thin accretion disk with spiral shocks, the position
of which is constant regarding to the components of a double system. First
of all a set of line profiles arising in an accretion disk with spiral
shocks was calculated. Profiles were calculated for various orbital phases.
After this shock-free model parameters were fitted to a minimum of a
residual deviation of a ''new'' shock-free model profile from an ''old''
spiral one.

Self-similar accretion flow solutions for logarithmic spiral waves were
obtained analytically by Chakrabarti (1990). For calculations of line
profiles arising in an accretion disk with spiral shocks we used the method
of Chakrabarti \& Wiita (1993). It should be noted, that for a given number
of spiral shocks, a model is uniquely determined by providing the pitch 
angle of the logarithmic spirals $\theta$ and the adiabatic index of the flow 
$\gamma $. The free parameters which
we varied for this analysis of line profiles are the number of spiral
shocks, spiral angle $\theta $ and azimuthal velocity component at the sonic
surface q$_{2c}$ (Chakrabarti \& Wiita 1993).
We would like to point that the model parameters (first of
all, brightness of shocks) were optimised to reproduce the real observational
emission line profiles in the spectra of dwarf novae.

To compute the shock-free model profile we used the method of Horne \&
Marsh (1986). Main free model parameters are $\alpha $, R and V. For details
see Horne \& Marsh (1986).
% and Horne (1995).

\begin{figure*}
%\picplace{8 cm}

{\vbox{
%\psfig{figure=fig4.ps,width=15cm,bbllx=0pt,bblly=0pt,bburx=370pt,bbury=200pt,clip=}}%
%}
\psfig{figure=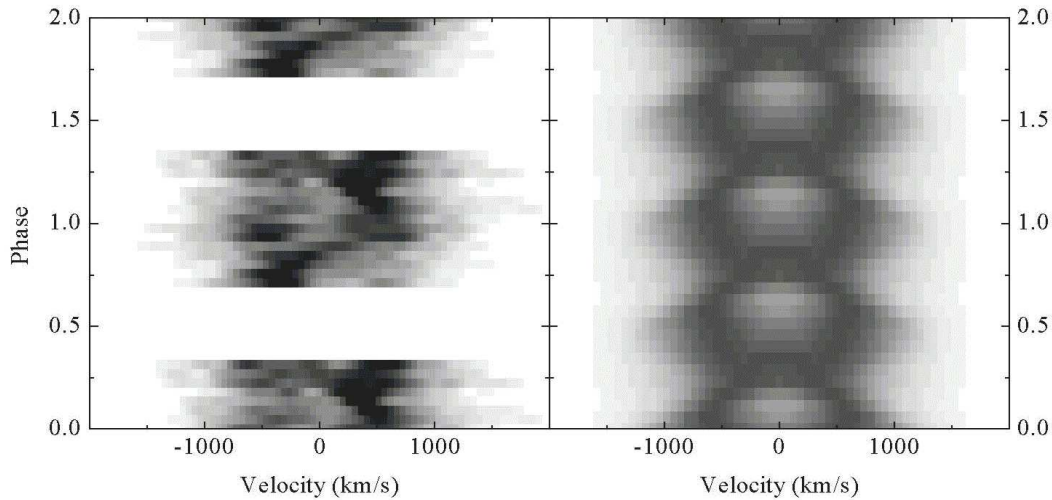,width=15cm,bbllx=43pt,bblly=259pt,bburx=568pt,bbury=531pt,clip=}}%
}
\par
\vspace{-1 cm}

\caption{
The observed (left panel) and the predicted without the influence of
hot spot (right panel) H$_{\beta }$ line profiles of U Gem}
\label{fig4}
\end{figure*}

\section{Analysis and comparison with observations}

As calculations have shown, the choice of spiral shocks parameters allows
one to obtain amplitudes of the variation of the double peak separation from
zero up to hundreds km/s. In this study, least deformed line profiles were
obtained at the value of the spiral angle $\theta $ of about 50$^{\circ
}\div $60$^{\circ }$. In this case the line parameters depend on the phase
of orbital period as $\sin n\varphi $ for an even number of shocks, and as
$\sin 3\varphi $/2 for an odd number, where n is a number of the shocks

For the spiral angle $\theta $ less than 50$^{\circ }$ in certain phases the line
profile becomes the one-peak for strong waves, or additional peak(s) appears
between the original two, and the amplitude of the variation of the
double peak separation will increase. For the spiral angle $\theta $ more than
60$^{\circ }$ additional peaks on the line wings are formed and the amplitude
of the variation of the double peak separation will sharply reduce (Fig.~\ref
{fig2}). Parameter q$_{2c}$ does not affect the behaviour of the line
profiles during the orbital period (but not on profiles themselves). Once
again note that we tended to examine the line profiles close to those
observed.

One can reproduce the detected variability of the line parameters of U Gem
(Borisov \& Neustroev 1998a) using the model of the accretion disk with two
symmetrically located spiral shocks with the spiral angle $\theta $ about 60$%
^{\circ }$, q$_{2c}$=0.025 and $\gamma $=1.5. In Fig.~\ref{fig3}, we present
a possible spatial image of the accretion disk emissivity model and the
predicted orbital phase dependence of the double peak separation along with
actual observed data. Note that we obtained an extremely similar pattern
with one for IP Peg (Steeghs et al. 1997). Figure ~\ref{fig4}
shows the predicted without the influence of hot spot and the observed
H$_{\beta }$ line profiles of U Gem.

\section{Conclusion}

In this paper we present a study of the observable spectral manifestations
of spiral shocks and compare them with results of spectroscopic observations
of U Gem. The primary result of this study is that the orbital behaviour of
the hydrogen emission line profile in the spectrum of U Gem can be
reproduced by the simple model of the accretion disk with two symmetrically
located spiral shocks with the spiral angle $\theta $ of about 60$^{\circ }$%
. The best evidence of the presence of shocks is the orbital phase
dependence of the separation of the double peak emission line. Furthermore,
even weak shocks may be expected to be detected from analysis of the double
peak emission line. Finally, the choice of the spiral shocks parameters
allows one to obtain amplitudes in the variation of the double peak
separation from zero up to hundreds km/s. This is perhaps the first
conclusive evidence for spiral shocks in the accretion disk of cataclysmic
variables in quiescence.

\begin{acknowledgements}

We thank V. Komarova, A. Golden and O. van den Berg for their help in
preparation of this paper. 
We would like to acknowledge J.Mattei and the AAVSO observers for providing
us with the photometric history of U Gem.
%We acknowledge with thanks the AAVSO for providing data from the AAVSO 
%International Database, based on observations submitted to the AAVSO by variable 
%star observers worldwide.
%In this research, we have used, and acknowledge with thanks, data from the AAVSO 
%International Database, based on observations submitted to the AAVSO by variable 
%star observers worldwide.
The work was partially supported by the Russian
Foundation for Basic Research (grant 95-02-03691).

\end{acknowledgements}

\end{document}